# Versatile SPH Open Boundary Conditions for Multiphase Flows in Extreme Conditions

Zi-Yang Zhan & Zhen Chen [†]

[a] *Marine Numerical Experiment Center, State Key Laboratory of Ocean Engineering, Shanghai 200240, China*

[b] *School of Ocean and Civil Engineering, Shanghai Jiao Tong University, Shanghai 200240, China*

## Abstract

Multiphase flows in pipes or channels, especially those in extreme flow conditions such as large density ratios, high Reynolds numbers, etc., are of particular interests to engineering applications such as oil/gas industry. Although the Smoothed Particle Hydrodynamics (SPH) method has been demonstrated as a promising numerical solver for multiphase flow problems due to its Lagrangian nature, its application to complex channel flow may encounter additional issues such as the open boundary condition and numerical instability in the extreme flow state. The present work aims to establish a general SPH algorithm for accurate and stable simulation of complex multiphase flows in configurations with open boundaries. The general scheme of weakly compressible SPH is adopted with special treatments implemented to ease the numerical oscillation in the density discontinuity scenario, and the turbulent model is implemented for interpretations of extreme flow conditions in high Reynolds numbers. Then, the conventional open boundary condition is fine-tuned by two new algorithms to guarantee the numerical stability in the inflow and the outflow regions. Firstly, a density relaxation is proposed to alleviate the pressure instability in the inflow region, which improves the smoothness of the particle pre-processing procedure. Secondly, the particle shifting technique with adaptive damper is implemented to adjust the magnitude of correction in the outflow region, which helps to suppress the velocity oscillation near the outlet. Validations of the proposed algorithms are carried out through four classic numerical examples, presenting appealing agreements with the analytical solutions and thus demonstrating the versatility in various flow conditions. Then, the robustness of the method is established through the turbulent multiphase channel flow cases and the horizontal slug channel flow with large density ratios. These results shed light on the value of the proposed algorithm as a general solver for complex multiphase flow problems with open boundaries.

[†] Correspondence author. Email address: zhen.chen@sjtu.edu.cn (Z. Chen)

# 1 Introduction

Multiphase flow with open boundary conditions represent a typical type of problems encountered in offshore and ocean engineering applications such as the transportation of oil and gas, submarine mining, and land reclamation by dredged spoil. Efforts towards better understanding of this type of flows would be favorable to the design and optimization of the devices and infrastructures deployed in these circumstances.

The initial effort made through a series of experiments[1-3] is essential to investigate the mechanism of complex channel flow. Meanwhile, considered as a promising complementary tool to offer comprehensive insights into the flow field, numerical simulation has been attracting growing interest in the study of complex channel flow[4-6]. However, characterized by the interfacial behavior with discontinuities in the physical field, the multiphase channel flow poses considerable difficulties to the conventional mesh-based methods. To deal with the existence of evolving interfaces, specific techniques like mesh-rezoning[7], level set (LS)[8], and volume of fluid (VOF)[9] are implemented to handle large interfacial deformations or to track the moving interface. In this aspect, with natural independence of the mentioned treatments, the Smoothed Particle Hydrodynamics (SPH)[10], as a meshfree particle method, has become an attractive solution for multiphase channel flow problems due to its Lagrangian nature[11].

Initially proposed for astrophysical problems[10,12], the SPH method has been extended to model various complex flow applications with moving boundaries and large deformation problems[13,14]. Efforts towards better computational accuracy and stability for SPH have been made through a series of algorithmic improvements, such as the pressure stabilization schemes[15-17] and the particle regularization techniques[18-20]. However, when it comes to the multiphase scene, the interrupted condition of the physical field still seriously destabilizes the simulation. In this regard, improved algorithms have been proposed for more stable simulation, including the modified discrete form of the governing equations [21-23], reformulation of the viscosity term [24] and the pressure gradient operator [25], revision of the diffusion scheme [26,27], refined surface-tension model[28-30], and the implementation of Riemann solver [31,32]. The mentioned algorithms have made considerable contributions in coping with the large density ratio condition and interfacial stability, but the existing simulations mainly focus on scenarios with simple boundaries, while the simulation for channel flows is still lacking.

Meanwhile, considerable efforts towards simulating the turbulent flows at high Reynolds numbers have been made through direct numerical simulation (DNS)[33] or implementing turbulence assumptions such as Reynolds-averaged Navier-Stokes (RANS) equations[34-36] and large eddy simulation (LES)[33,37] for the closure of the Navier-Stokes equation under the SPH framework. Since the former approach requires resolving a wide range of scales, the latter one becomes the more popular choice, which has been coupled to the SPH discretization with application in the simulation of various scenarios, including modeling of wave breaking[38], complex free surface flows[39,40], and the multiphase shallow water sloshing[41]. The above research provides a valuable reference for the coupling of turbulence models in SPH, but the simulation of complex flows in the channel is still to be explored.

To simulate the complex flows in the channel, it is an important challenge to incorporate the above algorithms into the specific boundary conditions of the channel. Particularly, the Lagrangian nature of SPH entails extra difficulties when dealing with open boundary conditions. To achieve the inflow-outflow condition without kernel function truncation, a feasible method is employing the semi-analytical boundary conditions[42], but additional boundary terms need to be introduced into the governing equations. On the contrary, another typical approach is to arrange in/outflow particles[43,44] at the upstream or downstream of the computational domain. In this case, the circulation of the flow is realized through the generation and removal of in/outflow particles. To suppress the fluctuations induced by the sudden jump of physical quantities from periodic particle generation/removal, several stabilization methods have been applied, including relaxation techniques [45] at the inlet region and the 'freezing treatments'[43,44] near the outlet boundary. Remarkable enhancements for the open boundary algorithm have been showcased in these works, and the extension to the multi-phase scenario has been realized[46,47]. However, it is worth noting that existing studies with the channel flow mostly absent the consideration of turbulence effects. The strategy for matching the turbulent state remains unspecified and thus calls for appropriate treatments to deal with the extra instabilities induced by the implementation of turbulent assumptions.

The present work aims to establish a general algorithm for accurate and stable simulation of complex channel flow, which can be adapted from single-phase to multi-phase scenarios with optional density ratios, involving the consideration of both laminar and turbulent states. In the present work, a WCSPH multiphase model is adopted to ensure the flexibility of the algorithm, whose versatility is then improved by the implementation of the turbulent model. To alleviate the numerical instability in extreme conditions, the conventional open boundary condition is fine-tuned by novel treatments. Firstly, the inflow region is divided into two sets of particles for a "two-step" relaxation to achieve a smooth preprocessing, and the pressure instability is addressed by the density relaxation treatment, which is applicable in both single-phase and multi-phase scenarios. Secondly, the particle regularization technique with an adaptive form is implemented to convey a more stable recovery of channel flows. Special treatment tailored for the outflow region is applied to the PST vector, with which the velocity oscillation induced by the outlet boundary could be prevented.

The remainder of this paper is arranged as follows. In section 2, the governing equations and the SPH discretization are given. Section 3 details the improved open boundary formulations of the proposed algorithm, where the density relaxation and the modified outflow PST vector are discussed. In section 4, four representative benchmark cases, including single-phase Poiseuille flow, two-fluid Poiseuille flow, immiscible two-phase co-current flow, and the 2D turbulent channel flow are simulated by the proposed method to validate its accuracy and convergence. The turbulent multiphase channel flow problems and the horizontal slug flow in the channel are resolved by the proposed algorithm to showcase its versatility in section 5, in which the robustness of the proposed algorithm is established through these two complex channel flow cases with high-density ratios. Conclusions are finally drawn in Section 6.

## 2 Governing Equations & SPH Discretization

The construction of the multiphase system is based on the physical recognition that both phases follow the conservation laws of mass and momentum, which can be expressed in the following Lagrangian form

$$\begin{cases} \frac{d\rho}{dt} = -\rho \nabla \cdot \boldsymbol{u} \\ \frac{d\boldsymbol{u}}{dt} = -\frac{\nabla p}{\rho} + \nu \nabla^2 \boldsymbol{u} + \boldsymbol{F} \end{cases} \tag{1}$$

where $\rho$, $p$, $\boldsymbol{u}$, $\nu$, $\boldsymbol{F}$ denote the density, the pressure, the velocity vector, the kinematic viscosity, and the external force, respectively.

In the weakly compressible SPH (WCSPH) framework, the assumption of weak compressibility allows explicit resolution of the pressure through the equation of state (EoS) [13] for the system closure. In this work, the Tait's EoS is used [48]

$$p = c_s{}^2(\rho - \rho_0) + p_0 \tag{2}$$

where $\rho_0$ is the initial reference density, $p_0$ is the background pressure applied to avoid negative pressures which may induce numerical instability, and $c_s$ is the artificial sound speed whose value should fulfill the following criterion of weak compressibility

$$c_s \geq 10 \, max\left( \sqrt{\frac{p_{max}}{\rho_0}} \, , u_{max} \right) \tag{3}$$

where $p_{max}$ and $u_{max}$ are the maximum predicted pressure and velocity, respectively. Additionally, in the present study, the same value of $c_s$ is applied to both phases for higher computational efficiency.

## 2.1 WCSPH model for multiphase flows

The SPH discretization consists of the kernel approximation and the particle approximation. A function $f(\boldsymbol{x})$ or its spatial derivative $\nabla \cdot f(\boldsymbol{x})$ can be written into the integral form and then replaced by the summation of the neighboring particles in the support domain, which gives

$$\langle f(\boldsymbol{x})\rangle_i = \int_\Omega f(\boldsymbol{x}') \; W(\boldsymbol{x} - \boldsymbol{x}') \; d\boldsymbol{x}' \approx \sum_j f(\boldsymbol{x}_j) W_{ij} V_j \tag{4}$$

$$\langle \nabla \cdot f(\boldsymbol{x})\rangle_i = \int_\Omega \nabla \cdot f(\boldsymbol{x}') \; W(\boldsymbol{x} - \boldsymbol{x}') \; d\boldsymbol{x}' \approx \sum_j f(\boldsymbol{x}_j) \cdot \nabla_i W_{ij} V_j \tag{5}$$

where $W$ is the kernel function; the subscript $i$ denotes the central particle; and $j$ refers to the neighboring particles within the support domain $\Omega$; $V_j$ is the volume of the particle $j$; $\boldsymbol{x}$ is the position vector of the particle. In the present study, the Gaussian kernel function is used.

Based on the SPH formulation in Eq.(4) & Eq.(5), the governing equations for multiphase system can be discretized into the following form:

$$\frac{d\rho_i}{dt} = -\rho_i \sum_j (\boldsymbol{u}_j - \boldsymbol{u}_i) \cdot \nabla_i W_{ij} V_j \tag{6}$$

$$\frac{d\boldsymbol{u}_i}{dt} = -\sum_j \left(\frac{p_i + p_j}{\rho_i \rho_j}\right) \nabla_i W_{ij} m_j + \frac{4\nu_i \nu_j}{\nu_i + \nu_j} \sum_j \frac{\boldsymbol{x}_{ij} \nabla_i W_{ij}}{(\boldsymbol{x}_{ij})^2 + 0.01 \bar{h}_{ij}^2} \boldsymbol{u}_{ij} V_j + \boldsymbol{F}_i \tag{7}$$

where the viscous term is written in the geometric mean form as suggested in Ref.[22] and other parameters used in Eq. are given as follows:

$$\boldsymbol{x}_{ij} = \boldsymbol{x}_j - \boldsymbol{x}_i, \boldsymbol{u}_{ij} = \boldsymbol{u}_j - \boldsymbol{u}_i, \bar{h}_{ij} = \frac{h_i + h_j}{2} \tag{8}$$

Although benefiting from the computational efficiency due to the explicit decoupling approach, WCSPH usually suffers from pressure instability. To alleviate this issue, the well-known δ-SPH [14,16] applies the diffusive term in the continuity equation to suppress the pressure oscillations, whose scheme is nurtured by Ref.[26] to address the unphysical oscillations that challenge the initial scheme in the multiphase scenario:

$$\frac{d\rho_i}{dt} = -\rho_i \sum_j (\boldsymbol{u}_j - \boldsymbol{u}_i) \cdot \nabla_i W_{ij} V_j + \delta h c_0 \sum_j \boldsymbol{\psi}_{ij} \cdot \nabla_i W_{ij} V_j \tag{9}$$

$$\boldsymbol{\psi}_{ij} = 2\left[\left(\delta\rho_j - \delta\rho_i\right) - \frac{1}{2}\left(\langle\nabla\delta\rho\rangle_i^L + \langle\nabla\delta\rho\rangle_j^L\right)\cdot\boldsymbol{x}_{ij}\right]\frac{x_{ij}}{\left(x_{ij}\right)^2 + 0.01h^2} \tag{10}$$

$$\delta\rho_i = \frac{P_i - P_0}{c_i^2} \tag{11}$$

$$\langle\nabla\delta\rho\rangle_j^L = \sum_j\left(\delta\rho_j - \delta\rho_i\right)\boldsymbol{L}_i\nabla_iW_{ij}V_j \tag{12}$$

where the coefficient $\delta$ is set as 0.1, and $\boldsymbol{L}_i$ is the kernel gradient correction (KGC) [49] matrix in the following form

$$\boldsymbol{L}_i = \left(\sum_j\left(\boldsymbol{x}_j - \boldsymbol{x}_i\right)\otimes\nabla_iW_{ij}V_j\right)^{-1} \tag{13}$$

Noting that Eq.(10) can be written as the initial δ-SPH scheme when simulating the single phase flow, since the reference density is the same for each pair of particles. Additionally, the cut-off value for particle density[25] is employed to prevent the instability induced by the negative pressure, and it should be emphasized that no artificial viscosity is adopted in our simulations.

## 2.2 SPH-based turbulent model

Considering the simulation of turbulent flows, two typical turbulent assumptions are adopted in the present study for the closure of the governing equations: the large eddy simulation (LES) and the Reynolds-averaged Navier-Stokes (RANS) equations.

For the former one, the momentum equation is modified by imposing the vortex viscosity term, which can be rewritten as follows:

$$\frac{d\boldsymbol{u}_i}{dt} = -\sum_j\left(\frac{p_i + p_j}{\rho_i\rho_j}\right)\nabla_iW_{ij}m_j + \frac{4\nu_i\nu_j}{\nu_i + \nu_j}\sum_j\frac{x_{ij}\nabla_iW_{ij}}{\left(x_{ij}\right)^2 + 0.01\hat{h}_{ij}^2}\boldsymbol{u}_{ij}V_j + \boldsymbol{\Theta}_i^T + \boldsymbol{F}_i \tag{14}$$

$$\boldsymbol{\Theta}_i^T = \frac{4\nu_i^T\nu_j^T}{\nu_i^T + \nu_j^T}\sum_j\frac{x_{ij}\nabla_iW_{ij}}{\left(x_{ij}\right)^2 + 0.01\hat{h}_{ij}^2}\boldsymbol{u}_{ij}V_j \tag{15}$$

Where $\nu_i^T$ is the turbulent viscosity, which can be calculated by:

$$\nu_i^T = \left(C_sl_f\right)^2\|\boldsymbol{D}\| \tag{16}$$

with

$$\|\boldsymbol{D}\| = \sqrt{2\boldsymbol{D}:\boldsymbol{D}} \tag{17}$$

where $C_s$ is the Smagorinsky constant with the value of 0.12 in the two-dimensional problems according to Ref.[38], and $l_f$ is the vortex filtering scale whose value is equal to the smoothing length in the present work. $\boldsymbol{D}$ is the strain rate tensor, which can be derived from:

$$\boldsymbol{D} = \frac{1}{2}\Big[\nabla\boldsymbol{u} + \big(\nabla\boldsymbol{u}\big)^T\Big] \tag{18}$$

For the latter one, the momentum equation is rewritten in a time-averaged form with the introduction of Reynolds stress, which depends on the time-averaged strain rate according to the Boussinesq hypothesis. To determine the unknowns in the rewritten momentum equation, the k-ε model is employed, whose expressions will be detailed in the appendix.

## 2.3 Solid boundary condition

The widely used Fixed Ghost Particle technique [14,50] is applied as the solid boundary treatment of the proposed method. Several layers of ghost particles are distributed along the solid walls. The pressure of the ghost particles is extrapolated from the known values in the inner particles with the Shepard algorithm, which gives:

$$p_i = \frac{\sum_j p_j W_{ij} V_j + \sum_j \rho_j (\boldsymbol{g} - \boldsymbol{a}_i)\cdot \boldsymbol{r}_{ij} W_{ij} V_j}{\sum_j W_{ij} V_j} \tag{19}$$

where $j$ refers to all the inner particles located in the support domain of the boundary particle $i$. The no-slip boundary condition on channel walls is realized by limiting the velocity of wall particles to 0.

## 2.4 Time marching scheme

The time marching in the present work is realized by the following modified prediction–correction scheme:

(1) Prediction step:

$$\begin{cases} \rho^i_{n+\frac{1}{2}} = \rho^i_n + \frac{\Delta t}{2}\left(\frac{d\rho}{dt}\right)^i_n \\ \boldsymbol{u}^i_{n+\frac{1}{2}} = \boldsymbol{u}^i_n + \frac{\Delta t}{2}\left(\frac{d\boldsymbol{u}}{dt}\right)^i_n \\ \boldsymbol{x}^i_{n+\frac{1}{2}} = \boldsymbol{x}^i_n + \frac{\Delta t}{2}\boldsymbol{u}^i_n + \delta\boldsymbol{x}^i_n \end{cases} \tag{20}$$

(2) Correction step:

$$\begin{cases} \rho^i_{n+1} = \rho^i_n + \Delta t\left(\frac{d\rho}{dt}\right)^i_{n+\frac{1}{2}} \\ \boldsymbol{u}^i_{n+1} = \boldsymbol{u}^i_n + \Delta t\left(\frac{d\boldsymbol{u}}{dt}\right)^i_{n+\frac{1}{2}} \\ \boldsymbol{x}^i_{n+1} = \boldsymbol{x}^i_n + \Delta t\boldsymbol{u}^i_{n+\frac{1}{2}} + \delta\boldsymbol{x}^i_{n+\frac{1}{2}} \end{cases} \tag{21}$$

The stability of the integration scheme is bounded by the Courant-Friedrichs-levy (CFL) condition [11], which leads to the following limit on the time step

$$\Delta t \leq CFL \cdot min\left(\frac{h}{c_s}, \frac{h^2}{\nu}, \sqrt{\frac{h}{|g|}}\right) \tag{22}$$

where $CFL$ is the Courant coefficient with a value of 0.125 in this study.

# 3 Improved Inflow/Outflow Conditions

To implement the inflow/outflow conditions without unphysical shock waves due to kernel truncation, a typical approach in the previous research is to arrange additional sets of particles at the entrance and the end of the flow, namely inflow particles and outflow particles. Such treatments successfully maintain the recycling of flow, but when it turns to extreme flow conditions such as a large density ratio or turbulent state, the numerical stability will be threatened. To address this issue, in the present study, the inflow particles will be divided into emitter particles and buffer particles (Fig. 1), and special treatments will also be implemented to ensure the stability of the inflow procedure.

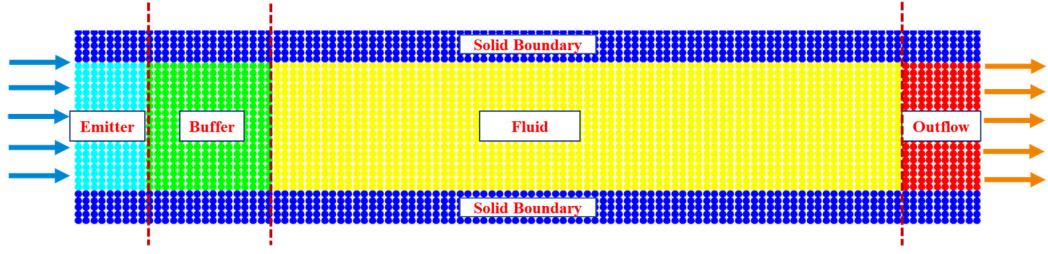

**Fig. 1. Schematic diagram of of particles classification.**

### 3.1 Relaxation strategy for inflow region

The existence of the open boundary at the inflow region affects the consistency of particle interpolation and thus induces pressure fluctuations, which cannot be sufficiently suppressed by the implementation of density diffusion. To cope with this, in the previous research, the inflow particles are suggested to be excluded from the SPH calculations with assigned parameters[43], but when the inflow particles flow into the fluid region, the sudden removal of the pressure restriction also affects the numerical stability. To this regard, instead of the exclusion treatment, we gradually increase the degrees of freedom of the inflow particles by involving them in the SPH governing equations with diminishing limitations, which achieves a smoother transition from the inflow region to the fluid region.

In this case, only the velocity is limited as the target value for the emitter particles, while a relaxation treatment will be imposed on pressure to suppress fluctuations:

$$\tilde{p}_i = \kappa p_{ref} + (1 - \kappa) p_i \tag{23}$$

where $\kappa$ is the relaxation factor. It can be assumed that the reference pressure $p_{ref}$ is approximate to the background pressure $p_0$ at the inflow region, and considering the equation of state in Eq.(2), the Eq.(23) can be rewritten as:

$$\tilde{p}_i = \kappa p_0 + (1 - \kappa)\left(c_s^2(\rho_i - \rho_{0,i}) + p_0\right) \tag{24}$$

Simplifying the above equation gives:

$$\tilde{p}_i = c_s^2\left((1 - \kappa)\rho_i + \kappa \rho_{0,i} - \rho_{0,i}\right) + p_0 \tag{25}$$

Employing the consistent form with Eq.(2), we can derive the following expression:

$$\tilde{p}_i = c_s{}^2(\tilde{\rho}_i - \rho_{0,i}) + p_0 \tag{26}$$

with

$$\tilde{\rho}_i = \kappa\rho_{0,i} + (1 - \kappa)\rho_i \tag{27}$$

Therefore, the pressure relaxation of the emitter particles can be transformed into a density-dominated form. In this case, the emitter particles can participate in the computation of the continuity equation with a density relaxation to suppress numerical instability. In addition, Eq.(27) is applicable in the multiphase scenarios without extra treatments, which preserves a generalized scheme for single-phase and multiphase flow.

For a higher degree of freedom at the downstream of the inflow region, once the emitter particles cross the threshold, they will be treated as buffer particles with the removal of pressure relaxation and the velocity assignment. To achieve improved numerical stability for the buffer particles, the velocity relaxation will be implemented [45]:

$$\widetilde{\boldsymbol{u}}_i = \epsilon\boldsymbol{u}_{ref,i} + (1 - \epsilon)\boldsymbol{u}_i \tag{28}$$

where $\boldsymbol{u}_{ref}$ is the target velocity of the stream. It should be noted that the velocity relaxation will be switched off as the buffer particles flow into the fluid region.

Noting that Eq.(27) and Eq.(28) can be rewritten in the following form:

$$\tilde{\rho}_i = \rho_i - \kappa(\rho_i - \rho_{0,i}) = \rho_i - \kappa\delta\rho_i \tag{29}$$

$$\widetilde{\boldsymbol{u}}_i = \boldsymbol{u}_i - \epsilon(\boldsymbol{u}_i - \boldsymbol{u}_{ref,i}) = \boldsymbol{u}_i - \epsilon\delta\boldsymbol{u}_i \tag{30}$$

which can be regarded as a damper to the density variation and the velocity variation, respectively. Therefore, both the mass conservation and the momentum conservation of the inflow region can be well preserved.

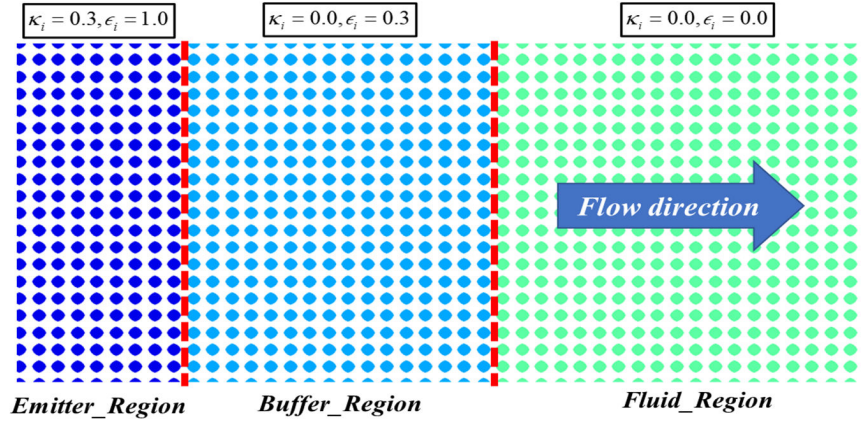

**Fig. 2. Schematic diagram of relaxation coefficients.**

Consequently, the relaxation strategy for the inflow region can be summarized as follows:

$$\begin{cases} \tilde{\rho}_i = \kappa_i \rho_{0,i} + (1-\kappa)\rho_i \\ \tilde{\boldsymbol{u}}_i = \epsilon_i \boldsymbol{u}_{ref,i} + (1-\epsilon_i)\boldsymbol{u}_i \end{cases} \tag{31}$$

The relaxation coefficients are defined with the suggested values (Fig. 2) as follows:

$$\begin{cases} \kappa_i = 0.3, \epsilon_i = 1.0 \ (i \in emitter\ region) \\ \kappa_i = 0.0, \epsilon_i = 0.3 \ (i \in buffer\ region) \\ \kappa_i = 0.0, \epsilon_i = 0.0 \ (i \notin inflow\ region) \end{cases} \tag{32}$$

Therefore, the inflow particles achieve a gradually increasing participation in the SPH calculation through the above relaxation treatment. Compared to the original method, improved numerical stability in the inflow region can be achieved through the present treatment, whose implementation has more simplicity by eliminating the necessity for extrapolation from the fluid domain.

### 3.2 Outflow-limiter for the generalized particle shifting technique

The Lagrangian nature of the SPH method may lead to unphysical concentrations of fluid particles, affecting the numerical stability. Several particle regularization techniques, including the particle shifting technique (PST) [19,20] and the transport-velocity formulation (TVF)[51,52], are proposed to alleviate this issue. In the present study, the PST is implemented to improve the regularity of particle distribution:

$$\delta \boldsymbol{x}_i = -CFL\frac{u_{max}}{c_0}(2h)^2 \sum_j \left[1 + 0.2\left(\frac{W_{ij}}{W(\delta_0)}\right)^2\right]V_j \nabla W_{ij} \tag{33}$$

where $\delta_0$ is the initial particle spacing.

Noting that the maximum velocity used in the above equation should be different for two phases in the multiphase system, the global maximum velocity $u_{max}$ is replaced by the following form:

$$u_{max}' = u_{max,1} \cdot \phi(i) + u_{max,2} \cdot \left(1 - \phi(i)\right) \tag{34}$$

where $u_{max,1}$ and $u_{max,2}$ are the maximum velocities of phase 1 and phase 2, respectively. The color function $\phi(i)$ is utilized to identify the particles from different phases, which can be expressed in the following form:

$$\phi(i) = \frac{\sum_j \phi_0(j)W_{ij}V_j}{\sum_j W_{ij}V_j} \tag{35}$$

$$\phi_0(j) = \begin{cases} 0 & phase\ 1 \\ 1 & phase\ 2 \end{cases} \tag{36}$$

Therefore, Eq.(34) allows flexible adjustment of the shifting magnitude and is thus more justifiable in a large density ratio scenario. It should be noticed that when calculating the single-phase flows, Eq.(34) can degrade to the initial form. Hence, the proposed treatment can be regarded as a more general form of PST for simulations of multiphase flows.

In addition, the PST is deeply dependent on the complete support domain of the kernel, whose effectiveness will be weakened by the truncation of the kernel function at the outflow region, and consequently leads to unphysical diffusion of particles. To address this issue, the component paralleling the flow direction of the PST vector is expected to vanish as the particle approaches the outflow boundary. In the present work, this expectation is fulfilled by introducing a diminishing factor to the specific component of the PST vector, which gives:

$$\delta \boldsymbol{x_i}^p = \delta \boldsymbol{x_i}^p \cdot \gamma \cdot \left( \frac{x_{max}^p - x_i^p}{L_{out}} \right)^2 \tag{37}$$

where $\gamma$ is the ratio of the length of the outflow region and the fluid region, $L_{out}$ is the length of the outflow region, and the superscript $p$ denotes the direction parallel to the flow. Such an outflow-limiter imposes a gradually increasing restriction on the shifting component normal to the outlet boundary, while preserving the tangential component to ensure the particle regularization. This modification to the outflow PST vector achieves a similar effect to the free-surface correction proposed in Ref.[20], while no additional free-surface detection is required, which expects higher computational efficiency.

## 4 Benchmark Validation

### 4.1 Single-phase Poiseuille flow

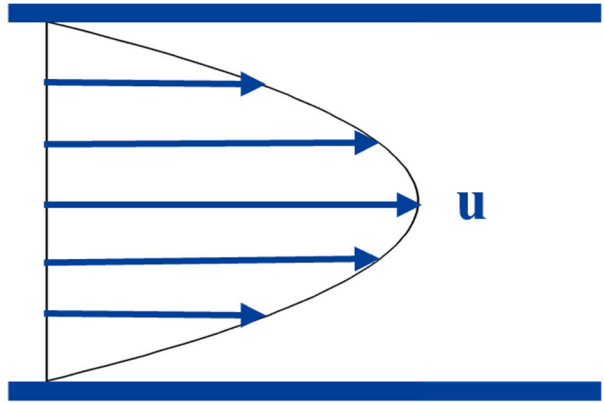

**Fig. 3. Schematic diagram of single-phase Poiseuille flow.**

The classic benchmark test of single-phase Poiseuille flow is resolved by the proposed algorithm to initially validate the accuracy of the methodology. As illustrated in Fig. 3, the liquid flows along the horizon direction of a channel with a diameter of D = 1mm. The density of liquid is $\rho$ = 1000 kg/m$^3$ with a dynamic viscosity $\mu$ = 0.001m Pa•s. The flow is driven by the inflow stream velocity with the amplitude $U_{max}$ = 0.1mm/s in the horizon direction, and the no-slip boundary condition is employed in this case. Additionally, a uniform velocity is arranged for the initial condition of the channel, and the flow field evolves with the inflow stream during the simulation.

The velocity profiles obtained from the present SPH algorithm are compared with the analytical solution in Fig. 4. As shown in the figure, close agreement has been achieved with analytical solutions under three sets of numerical results with different particle spacings. For further evaluation, the convergence is examined in Fig. 5, where the local velocity profiles present a gradual convergence with the decreasing particle spacings. The time evolution of the L2 error between the present SPH method and the analytic solution is shown in Fig. 5(b), from which it can be observed that the flow field is gradually updated by the inflow condition and consequently evolves to a stable state. The numerical errors also reduce when the spatial resolutions increase, which initially demonstrates the convergence of the present simulation.

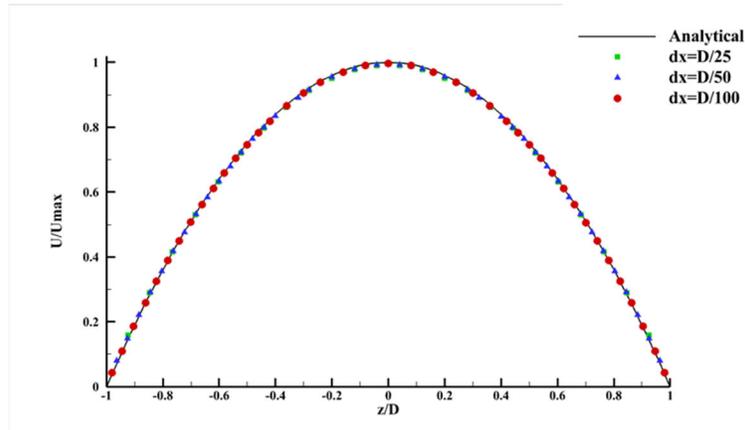

**Fig. 4. Velocity profile obtained by the present SPH method in comparison with the analytical solution.**

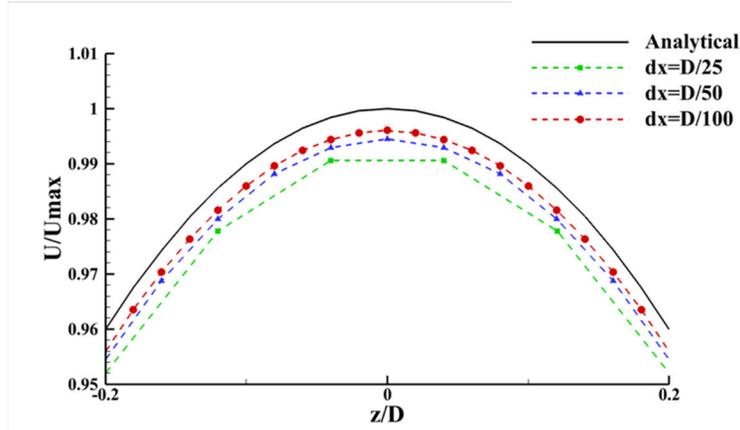

**(a) Local view of velocity profiles obtained by the present SPH method in comparison with the analytical solution.**

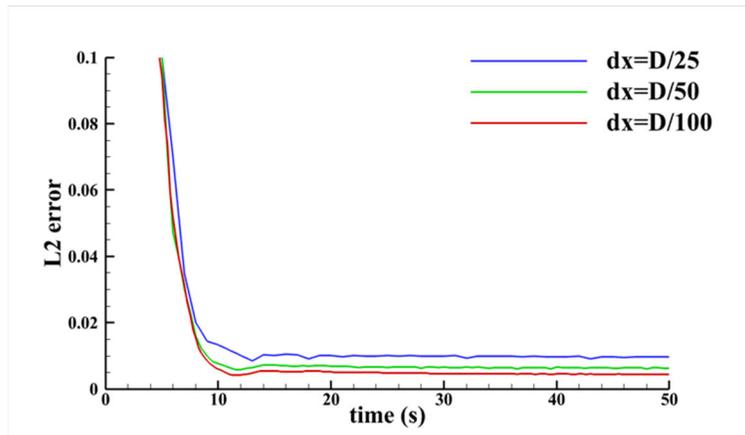

**(b) Time evolution of the L2 error between the present SPH method and the analytic solution**

**Fig. 5. Convergence test of the single-phase Poiseuille flow**

## 4.2 Two-fluid Poiseuille flow

With sensitivity to viscosity models, the two-fluid Poiseuille flow is a typical benchmark to validate the multiphase scheme. In this subsection, we use this example to validate the effectiveness of the multi-phase model employed in the proposed algorithm.

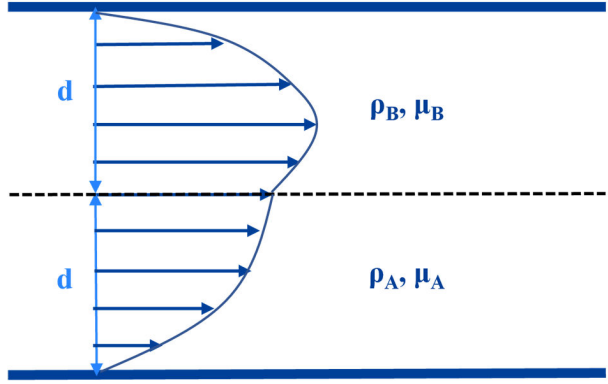

**Fig. 6. Schematic diagram of two-fluid Poiseuille flow.**

The arrangement of this problem is given in Fig. 6, where two fluids are flowing in a channel with a diameter of D = 2d =1mm. The top half of the channel is filled with fluid B, whose density is $\rho_B$ = 100 kg/m$^3$ with a dynamic viscosity $\mu_B$ = 0.001Pa•s, while the other half is filled with fluid A, with $\rho_A$ = 1000 kg/m$^3$ and $\mu_A$ = 0.004 Pa•s. The general analytical solution for this problem is derived in Ref.[53], which reads:

$$\boldsymbol{u}_A = F_x \frac{d^2}{2\mu_A}\left[\left(\frac{2\mu_A}{\mu_A+\mu_B}\right) + \left(\frac{\mu_A-\mu_B}{\mu_A+\mu_B}\right)\left(\frac{z}{d}\right) - \left(\frac{z}{d}\right)^2\right] \tag{38}$$

where the driving force per unit mass is defined as $F_x$ = 0.2m/s$^2$. A similar formulation can be derived to describe fluid B, and the no-slip boundary condition is employed for the horizontal boundary in this case. The velocity profiles obtained from the present algorithm are compared with the analytical solutions in Fig. 7(a), where $u_{avg}$ is the average velocity of the channel. A good agreement can be observed in Fig. 7(a) with different spatial resolutions. The convergence analysis is also given in Fig. 7(b), where the L2 norm of error is obtained from four spatial resolutions, including dx = D/12.5, dx = D/25, dx = D/50, and dx = D/100. It can be observed that the error has a superlinear decrease with the increasing particle resolution, thus the present algorithm shows a convergence approximately larger than the first order.

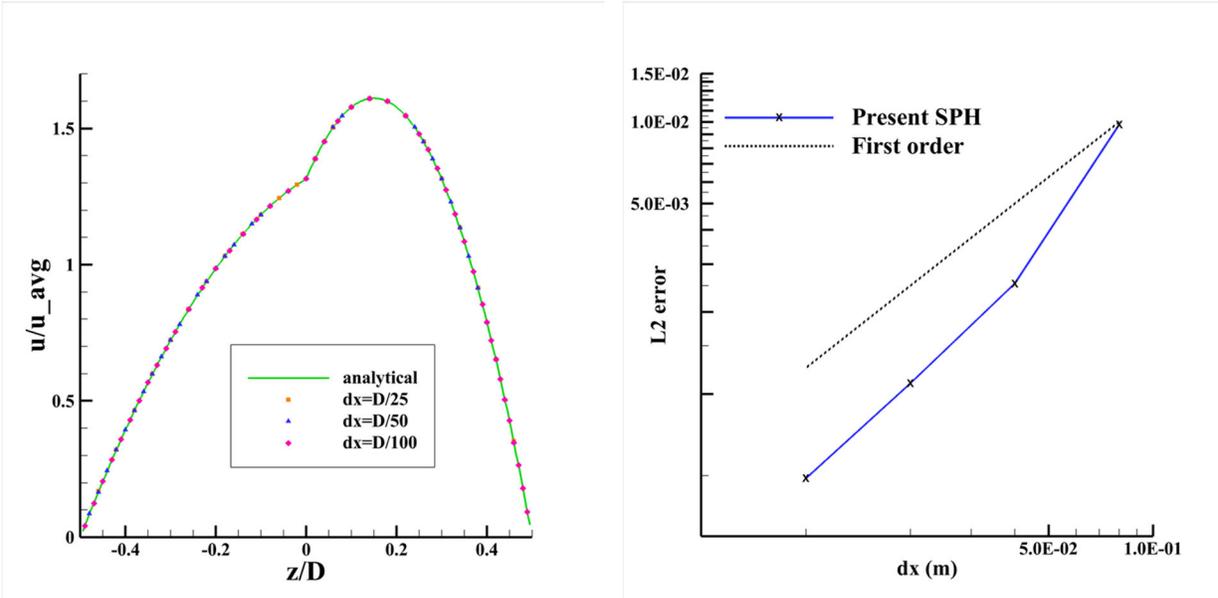

**(a)** Velocity profile obtained by the present SPH method in comparison with the analytical solution[53]

**(b)** Convergence analysis

**Fig. 7. Analysis of two-fluid Poiseuille flow**

## 4.3 Immiscible two-phase co-current flow

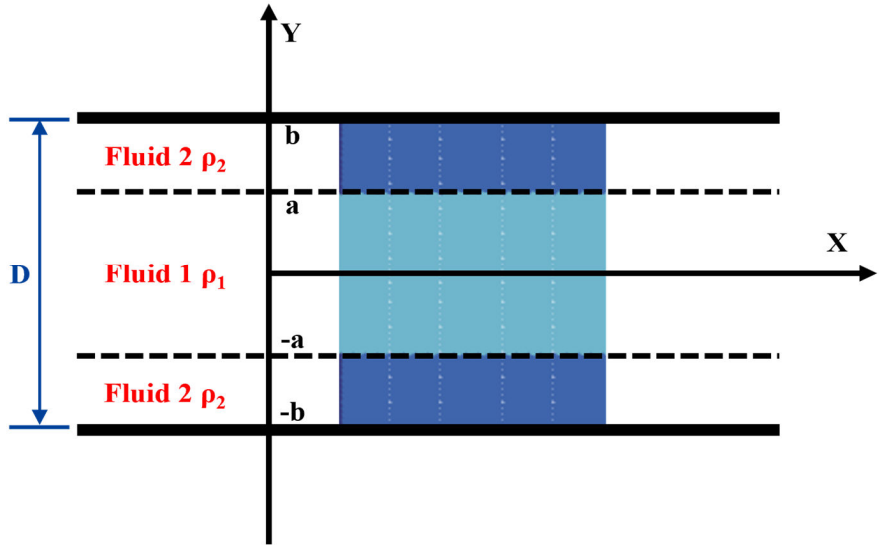

**Fig. 8. Schematic diagram of immiscible two-phase co-current flow.**

Since the density does not affect the solutions in the two-fluid Poiseuille flow case, to further evaluate the accuracy of the proposed algorithm in multiphase problems with significant density ratios, the benchmark test of immiscible two-phase co-current flow is carried out in this section. The physical configuration of this problem is shown in Fig. 8, where the region $|y| < a$ of the channel is filled with the fluid 1 of density $\rho_1$ while fluid 2 of density $\rho_2$ occupies the rest region within the channel. Following the arrangement in Ref, the density of fluid 2 is $\rho_2 = 1$ with different density ratios of $\rho_2/\rho_1$, the kinetic viscosities for the fluid are $v_1 = v_2 = 0.01$, and the parameters of the channel are set as: a=50, b=100. Additionally, an external force $F_x = 1 \times 10^{-9}$ is imposed on either Fluid 1 or Fluid 2 to drive the multiphase system, and the particle spacing is set as D/100, similar to the grid size in Ref.[54].

For the scenario of the external force applying on Fluid 1, the velocity profiles of the flow obtained by the present algorithm are presented in Fig. 9, in which the density ratios from 10 to 1000 are considered. It can be observed that the numerical results achieve good agreement with the analytical solutions given in Ref.[55], which gives initial validation of the accuracy of the proposed algorithm.

A comparison of the velocity profile is then shown in Fig. 10, with attention paid to the results obtained by the present algorithm with or without the density relaxation treatment. It can be observed that the implementation of the density relaxation remedies the issue of velocity excess caused by pressure fluctuations in the original counterpart. A similar comparison focusing on the PST outflow limiter is given in Fig. 11, which shows that the increase in velocity is well suppressed by applying the PST outflow limiter, which indicates the effectiveness of this treatment, and thus the mass and momentum conservation can be well preserved in the present algorithm.

The scenario of the external force applying on Fluid 2 is also simulated, whose results are presented in Fig. 12 and Fig. 13. It can be observed that good agreements have been achieved with different density ratios, but the velocities at the interfaces are slightly smaller in the present algorithm compared to the analytical solution. This appealing result further demonstrates the accuracy of the proposed algorithm in simulating high-density-ratio multiphase flow in a confined channel.

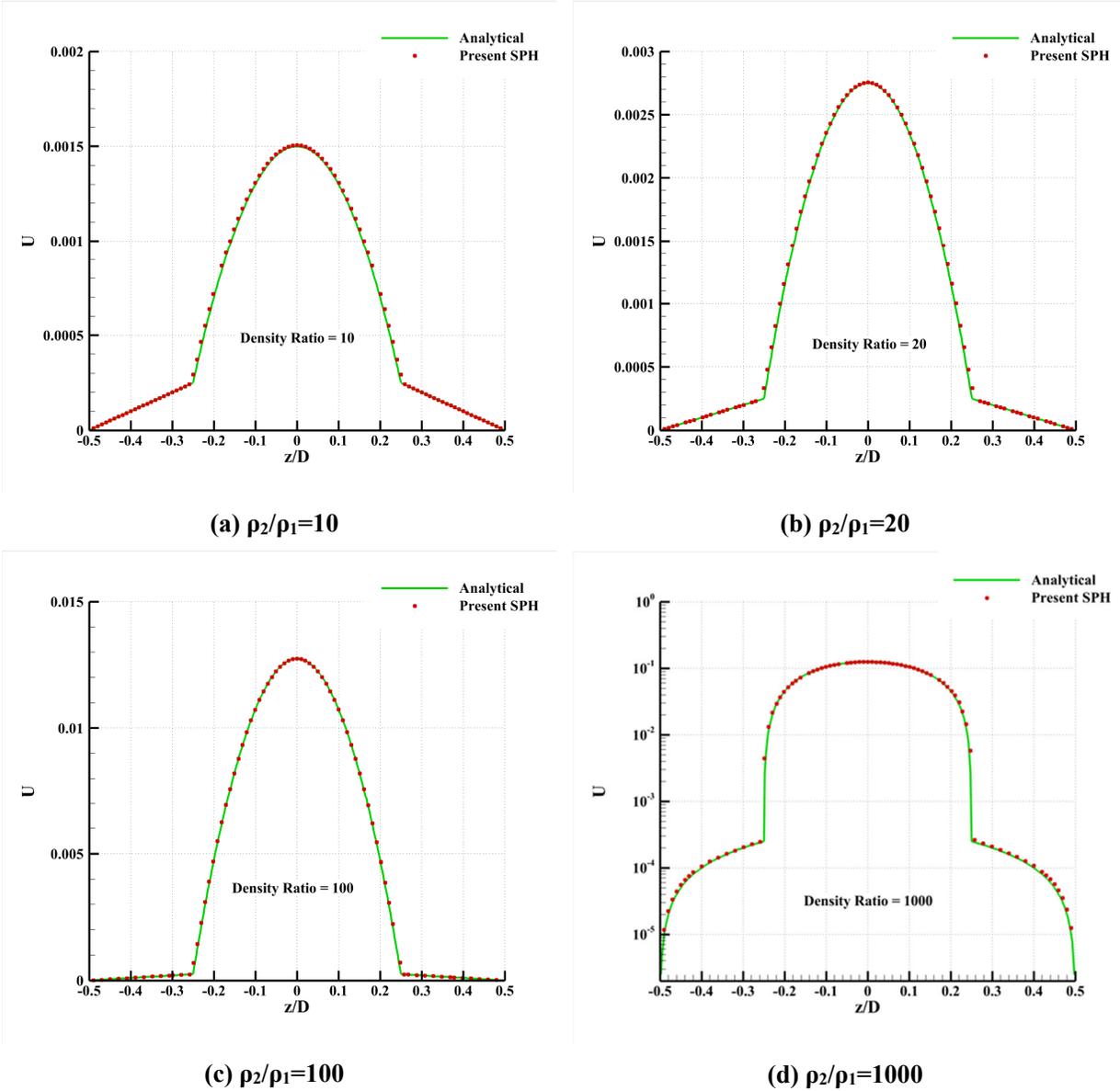

**(a) $\rho_2/\rho_1$=10**

**(b) $\rho_2/\rho_1$=20**

**(c) $\rho_2/\rho_1$=100**

**(d) $\rho_2/\rho_1$=1000**

**Fig. 9. Velocity profiles of immiscible two-phase co-current flow in comparison with the analytical solution[55] (External force on fluid 1)**
**(The velocity in (d) is shown in the log scale due to the extremely small value).**

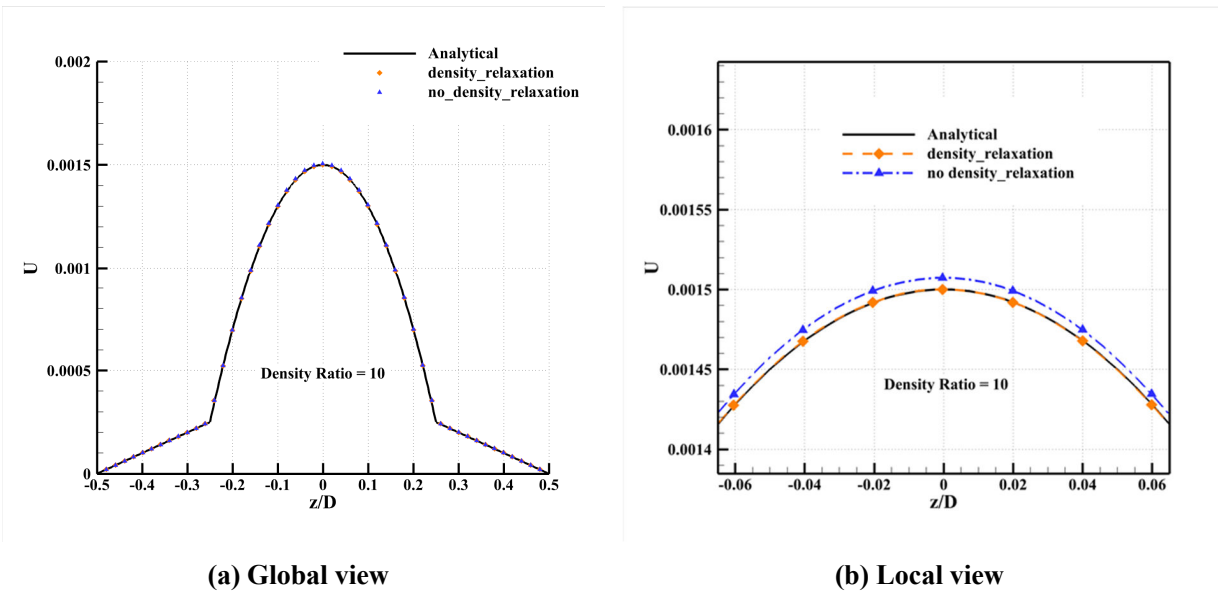

**(a) Global view**          **(b) Local view**

**Fig. 10. Comparison of the velocity profile with/without density relaxation for the case of immiscible two-phase co-current flow (External force on fluid 1)**

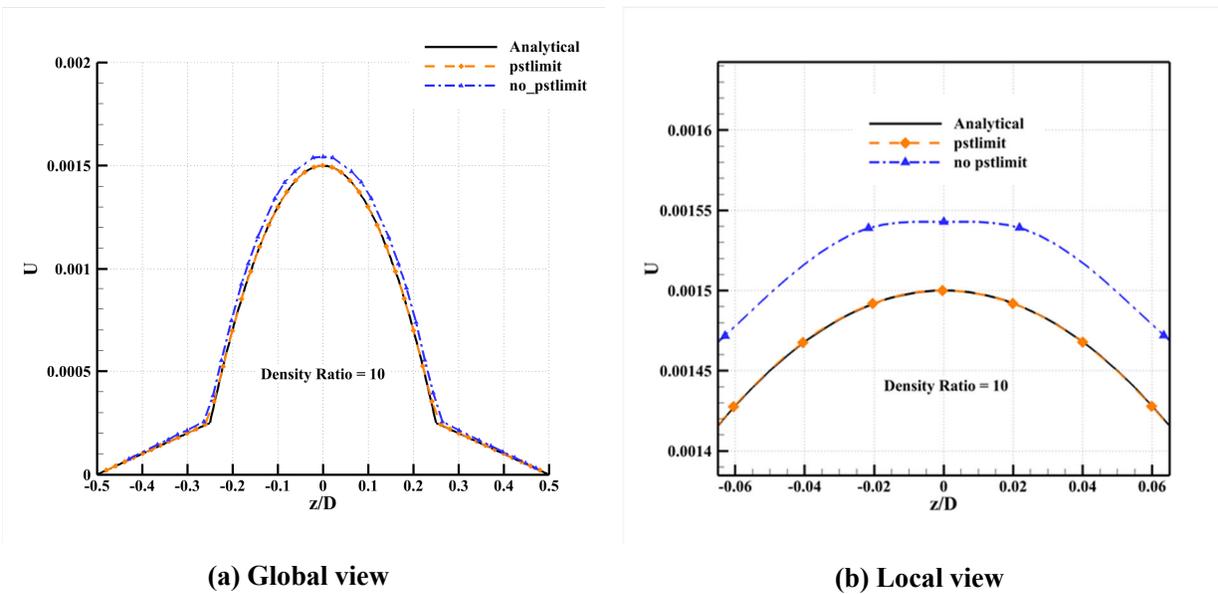

**(a) Global view**          **(b) Local view**

**Fig. 11. Comparison of the velocity profile with/without PST outflow limiter for the case of immiscible two-phase co-current flow (External force on fluid 1)**

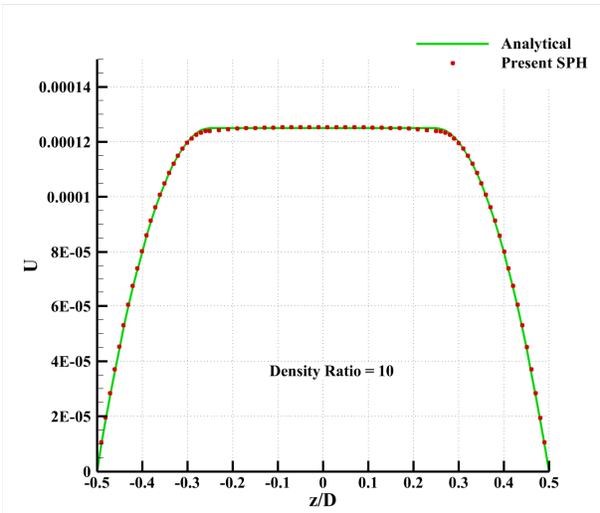

**(a) $\rho_2/\rho_1$=10**

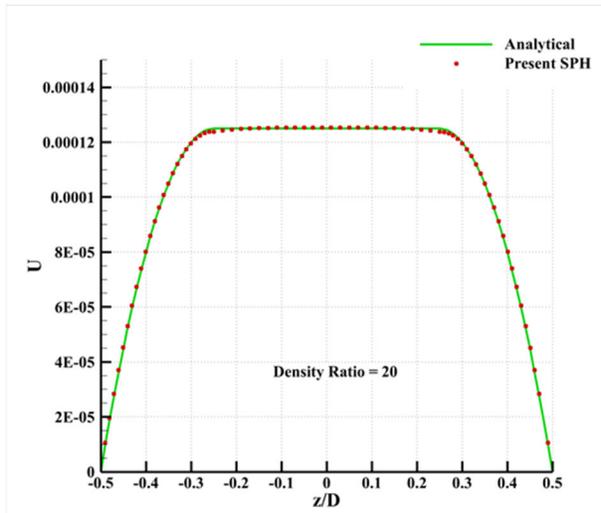

**(b) $\rho_2/\rho_1$=20**

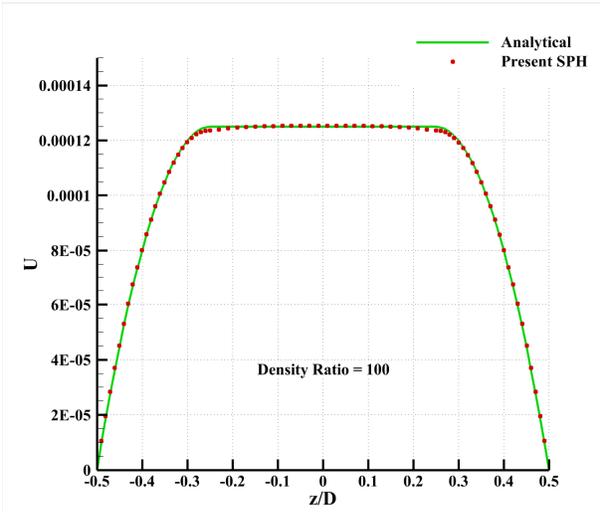

**(c) $\rho_2/\rho_1$=100**

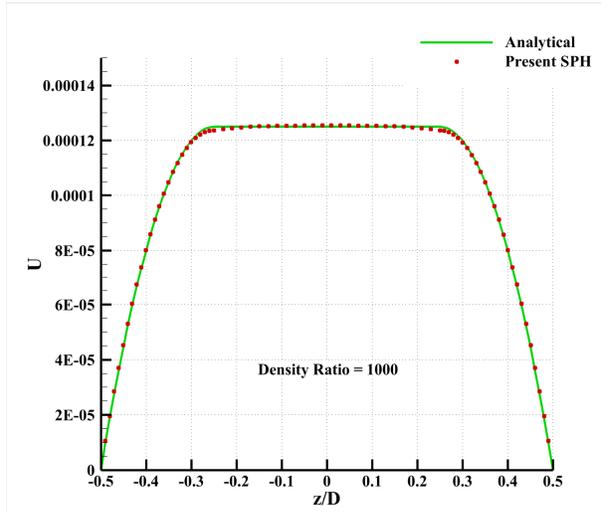

**(d) $\rho_2/\rho_1$=1000**

**Fig. 12. Velocity profiles of immiscible two-phase co-current flow in comparison with the analytical solution[55] (External force on fluid 2)**

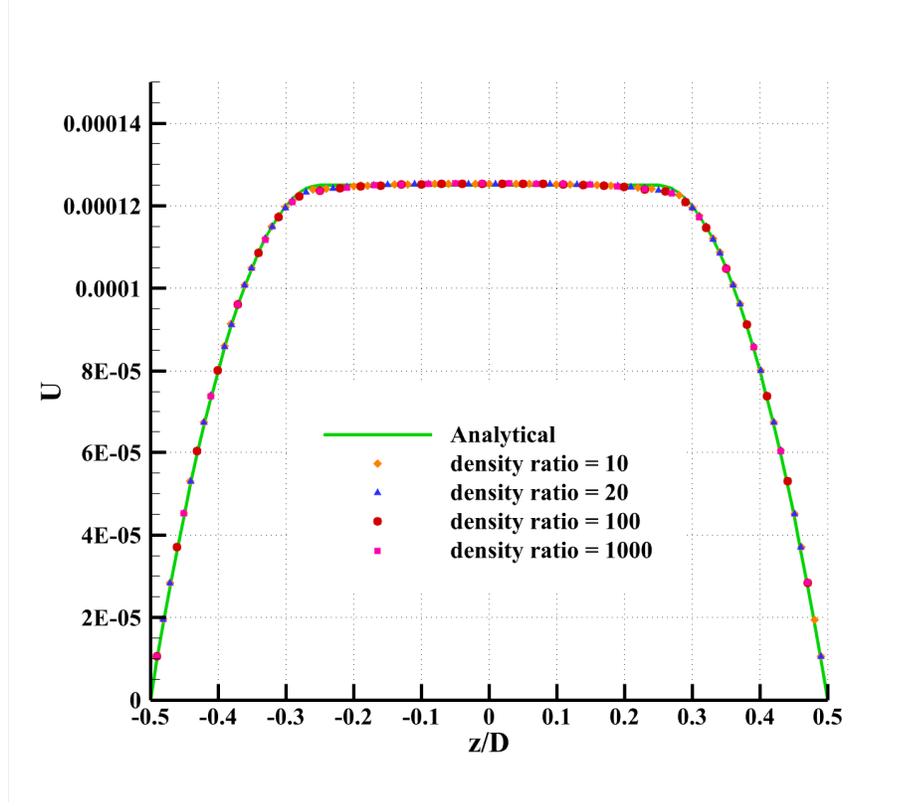

**Fig. 13. Velocity profiles of immiscible two-phase co-current flow in comparison with the analytical solution[55] (External force on fluid 2)**

### 4.4 2D turbulent flow in a channel

To validate the effectiveness of the proposed algorithm under a turbulent flow state, a 2D channel turbulent flow case is simulated. The configuration of this case is shown in Fig. 14, in which the length of channel L is 40m and the diameter D is set as 2 m. The no-slip conditions are applied on the solid walls, and the pressure at the outlet region is set as the background value. According to Ref.[56], the Reynolds number is set as 12,300, and the inlet flow velocity profile is described by the one-seventh power-law, whose expression is given as:

$$u(z) = \begin{cases} u_{max} \cdot \left(D - \dfrac{2z}{D}\right)^{\frac{1}{7}}, & z > 1.0 \\ u_{max} \cdot \left(\dfrac{2z}{D}\right)^{\frac{1}{7}}, & z \leq 1.0 \end{cases}, \tag{39}$$

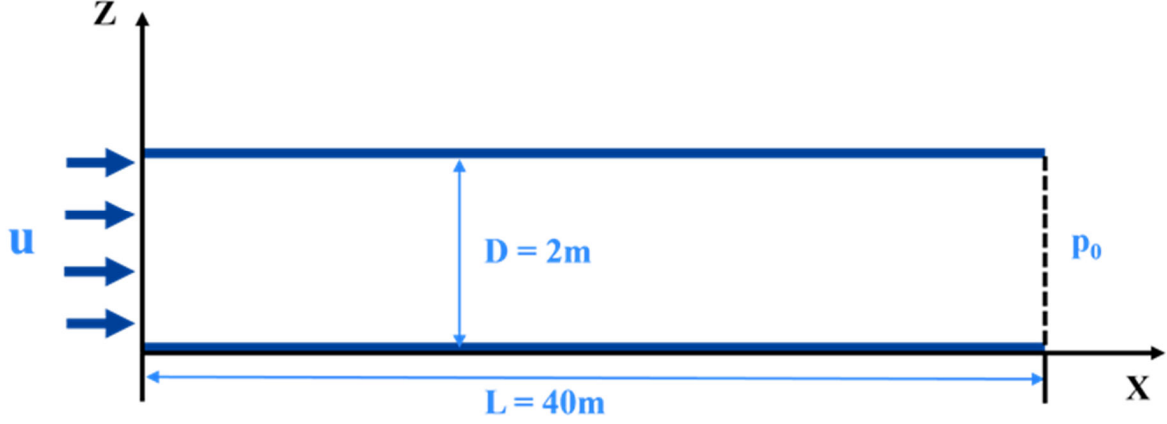

**Fig. 14. Schematic diagram of 2D turbulent flow in a channel**

where $u_{max}$ is the max horizontal velocity with the value of 1.23m/s, and the dynamic viscosity of the fluid is set as 0.0001Pa•s. Before evaluating the computational results, it is worth emphasizing that the treatment for the wall regions significantly affects the accuracy of turbulence prediction. Adaptive resolution is a proper way to resolve LES under the near-wall scales, which is similar to the refined mesh of the mesh-based method, while another approach recommended in the previous research is applying the wall functions[33]. In the present paper, since the variable resolutions have not been incorporated into the current algorithm, the wall function in ref.[56] is applied to model the flow evolving near the wall, whose definition is presented as follows:

$$\begin{cases} \frac{u}{u_\tau} = \frac{\rho u_\tau y}{\mu} \, , \quad y^+ \leq 11.225 \\ \frac{u}{u_\tau} = \frac{1}{\kappa} \ln \frac{\rho u_\tau y}{\mu} + B \, , \quad y^+ > 11.225 \end{cases} \cdot \tag{40}$$

where $u_\tau$ is the friction velocity, whose definition can be found in ref. $y^+ = \frac{\rho u_\tau y}{\mu}$ is the normalized distance to the wall, $\kappa = 0.4187$ and $B = 5.45$ are the Von Karman constant and the roughness constant, respectively. It is worth stressing that the wall function is only applied in the near-wall region, with the suggested threshold as $y^+ = 60$ in the present work.

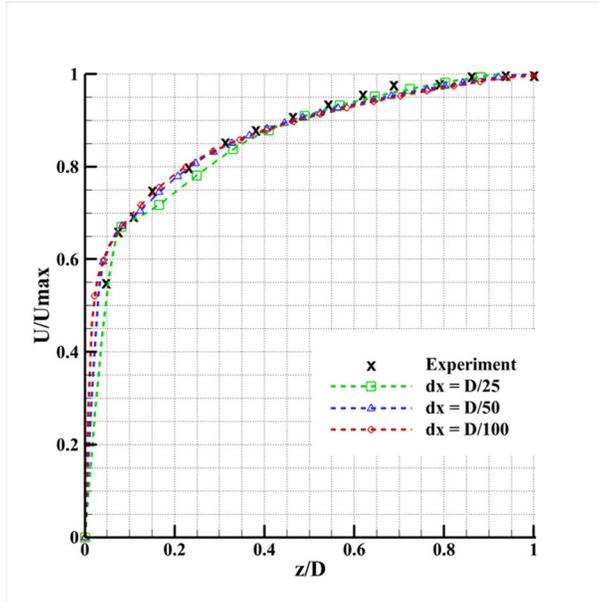

**Fig. 15. Velocity profiles of 2D turbulent channel flow in comparison with the experiment data[57] (LES model).**

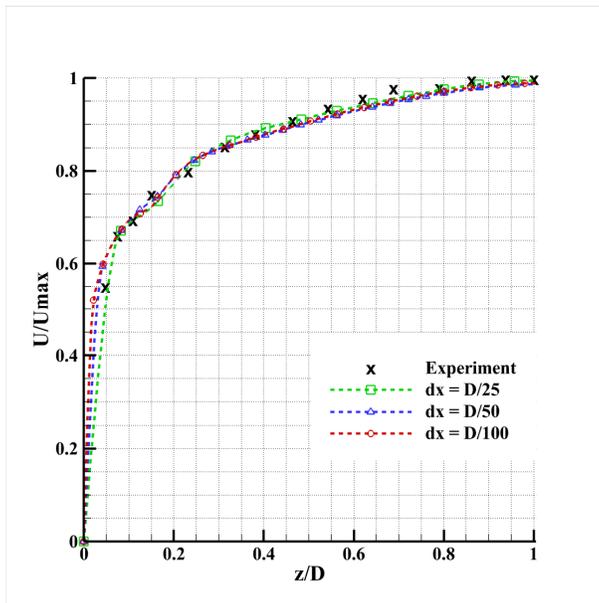

**Fig. 16. Velocity profiles of 2D turbulent channel flow in comparison with the experiment data[57] (k-ε model).**

The fully developed velocity profiles obtained by the present algorithm with the LES model are presented in Fig. 15 and compared with the experiment data[57], in which the convergence is examined by considering three sets of spatial resolutions dx = D/25, dx = D/50, and dx = D/100.

It can be observed that the velocity developments under different resolutions present a consistency, especially the difference to experimental data presents a significant reduction under the resolution of dx=D/50 compared to the lower resolution set, and the dx=D/100 set shows a good convergence to the experiment data. The results obtained by the k-ε model are also shown in Fig. 16 and present comparable accuracy and consistency with the experimental results. Thus, the above results obtained from the two turbulent models provide preliminary confirmation of the accuracy of the proposed method.

To validate the effectiveness of the density relaxation treatment in the turbulent state, the comparison of the velocity profile obtained by LES model is presented in Fig. 17. An anomalous enlargement of velocity can be observed in the original sets, while such an issue is addressed by the density relaxation with more consistent with the experimental data. Fig. 18 displays the results obtained from k-ε model, where the effect of suppressing the velocity enlargement by the density relaxation can be noted, which demonstrates an effective performance of density relaxation in improving the numerical stability for the turbulent problems.

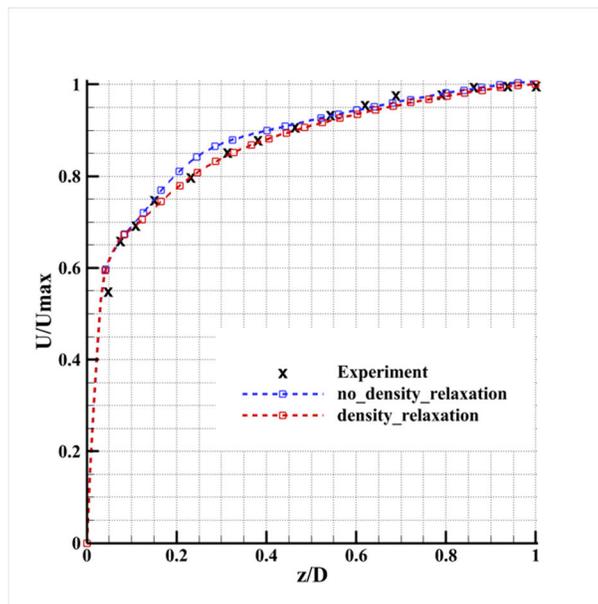

**Fig. 17. Comparison of the velocity profile with/without density relaxation for the case of 2D turbulent channel flow (LES model).**

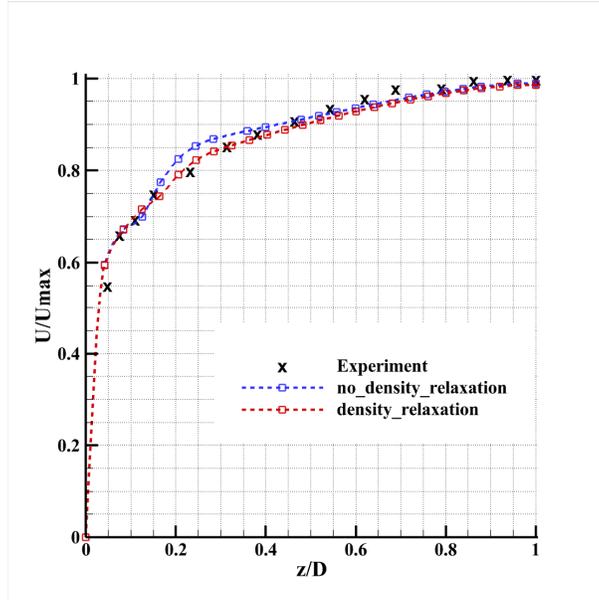

**Fig. 18. Comparison of the velocity profile with/without density relaxation for the case of 2D turbulent channel flow (k-ε model).**

# 5 Numerical examples

## 5.1 Turbulent multiphase channel flow

For a more comprehensive evaluation of the proposed algorithm, turbulent multiphase channel flows with different density ratios are simulated in this section, whose initial arrangement is shown in Fig. 19. Containing two fluids, the diameter of the channel is D = 0.2m with a 20D length, of which fluid 2 occupies the center of the space with a width of H. The density of fluid 1 is set as $\rho_1 = 1000$ kg/m$^3$ with the kinetic viscosity $v_1 = 1 \times 10^{-6}$ m$^2$/s, and fluid 2 has the same kinetic viscosity as the former while its density is determined by the density ratio of the specific case. A uniform inflow velocity $U_{in}$=2m/s is applied, which in conjunction with the mentioned physical properties determines the Reynolds number of the flow is 400,000. The background pressure is set as $p_{0=}\rho_1 U_{in}^2/2$, and it should be noted that the gravity effect is not taken into account.

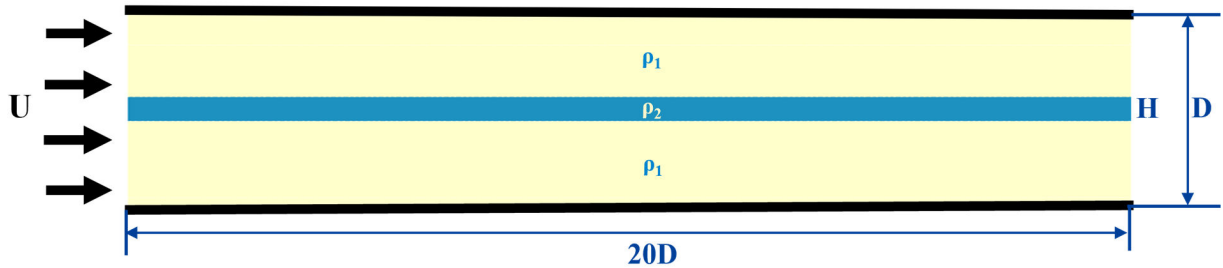

**Fig. 19. Schematic diagram of turbulent multiphase channel flow**

The major concern in the numerical simulation of multiphase channel flow in turbulent state is the stability of the computational progress. Thus, to justify the effectiveness of the proposed treatments, two sets of validations for the proposed algorithm are carried out with attention paid to the results obtained by the present algorithm with or without the corresponding treatment.

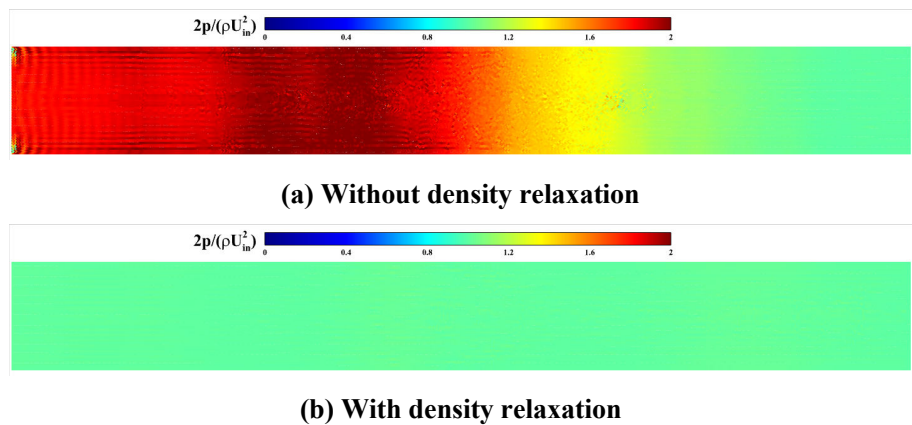

**(a) Without density relaxation**

**(b) With density relaxation**

**Fig. 20. Pressure field of the turbulent multiphase channel flow obtained by the present algorithm ($\rho 2/\rho 1 = 1/10$).**

Fig. 20 presents the effectiveness of the density relaxation algorithm in turbulent multiphase flow conditions. The comparison of the pressure field proves that the proposed algorithm can significantly suppress the numerical fluctuation. Fig. 21(a) and (b) depict a zoom-in view of the velocity field of the turbulent multiphase channel flow, which illustrates that the outflow-limiter of the PST remedies the anomalous surging of velocity observed in the original simulation.

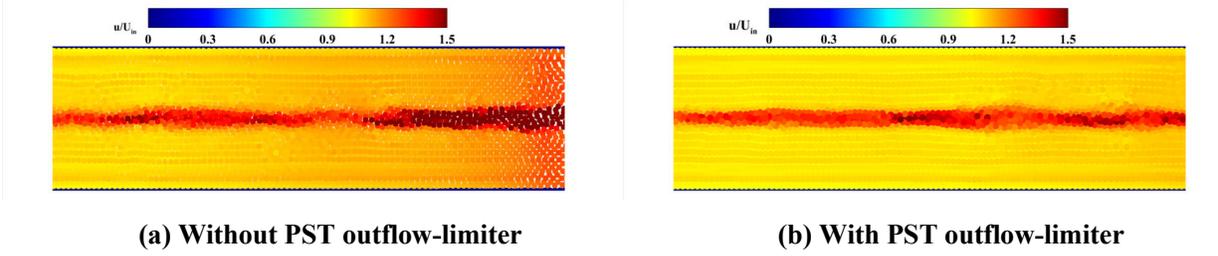

**(a) Without PST outflow-limiter**    **(b) With PST outflow-limiter**

**Fig. 21. Velocity field of the turbulent multiphase channel flow (at outflow region) obtained by the present algorithm (ρ2/ρ1=1/10).**

To confirm the robustness of the present method, a density ratio 1000 example is simulated for the convergence test, where the width of fluid 2 is set as H=0.1D,0.2D, and 0.3D respectively. To obtain the velocity profile, the velocity information carried by the particles at a specified profile is collected within a certain period, and then a sampling procedure is employed for the scatters. Therefore, the pre-validation is conducted to ensure duration and position convergence, respectively.

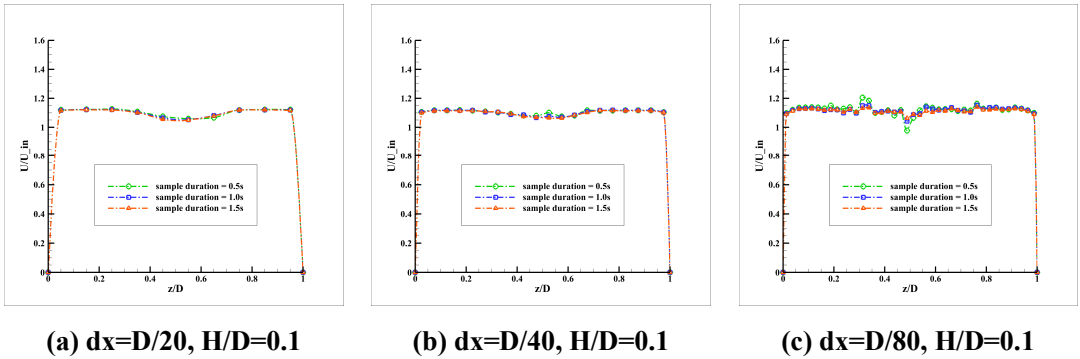

**(a) dx=D/20, H/D=0.1**    **(b) dx=D/40, H/D=0.1**    **(c) dx=D/80, H/D=0.1**

**Fig. 22. Pre-validation for the sample duration convergence.**

**(ρ2/ρ1=1/1000, x=10D)**

As shown in Fig. 22, the velocity profile converges when the sampling durations approach 1s and 1.5s, with less fluctuation compared to the 0.5s group. In this case, the sampling duration of 1.5s is adopted for the following tests. The convergence test for sampling position is presented in Fig. 23, where the results at the three chosen cross-sections show good consistency, which proves that the flow has a full development after the half of the channel, and the x=12.5D is used for further evaluation.

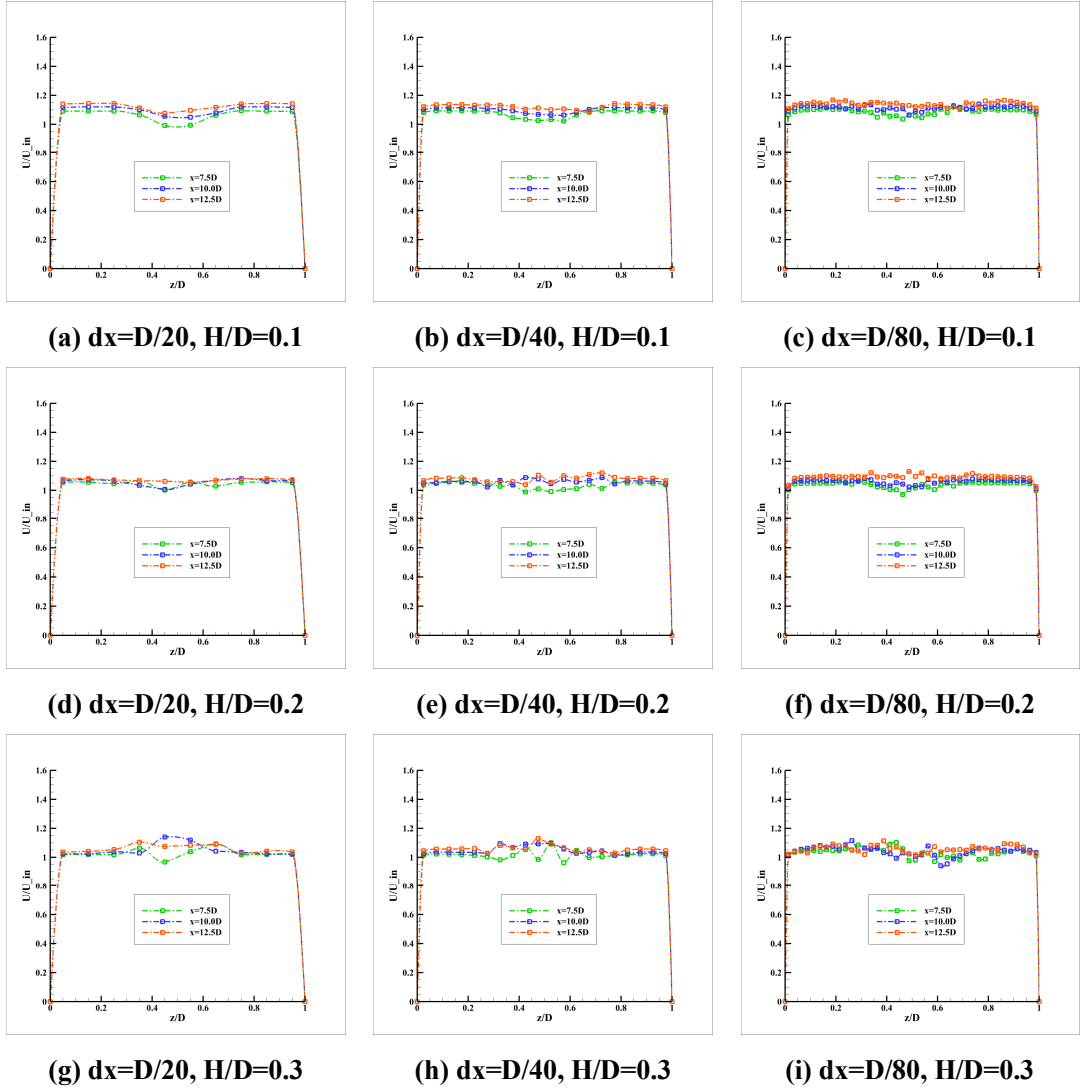

**(a) dx=D/20, H/D=0.1**    **(b) dx=D/40, H/D=0.1**    **(c) dx=D/80, H/D=0.1**

**(d) dx=D/20, H/D=0.2**    **(e) dx=D/40, H/D=0.2**    **(f) dx=D/80, H/D=0.2**

**(g) dx=D/20, H/D=0.3**    **(h) dx=D/40, H/D=0.3**    **(i) dx=D/80, H/D=0.3**

**Fig. 23. Pre-validation for the location convergence.**

**(ρ2/ρ1=1/1000)**

The convergence test for spatial resolution is performed in Fig. 24 and the results with different resolutions show good convergence, which demonstrates the stability of the present method, and the particle resolution of dx = D/80 is adopted in the following discussions.

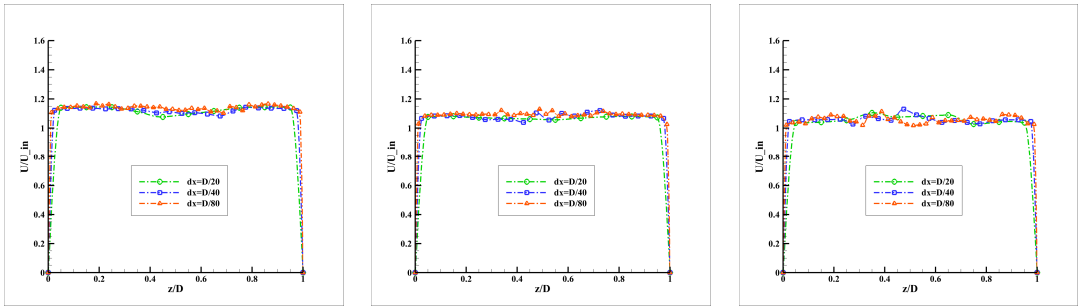

**(a) H/D=0.1**   **(b) H/D=0.2**   **(c) H/D=0.3**

**Fig. 24. Convergence test for the spatial resolution.**

**(ρ2/ρ1=1/1000)**

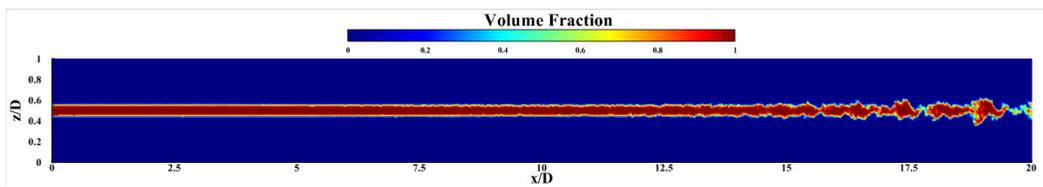

**(a) Distribution of volume fraction (ρ2/ρ1=1/10, H/D=0.1)**

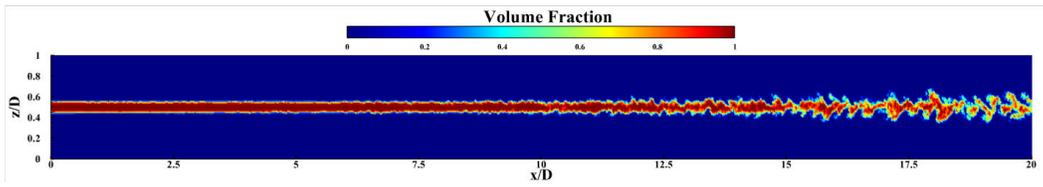

**(b) Distribution of volume fraction (ρ2/ρ1=1/100, H/D=0.1)**

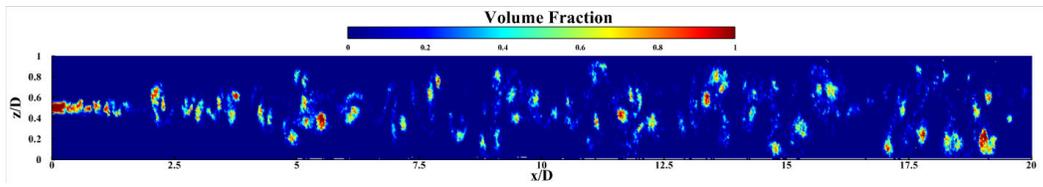

**(c) Distribution of volume fraction (ρ2/ρ1=1/1000, H/D=0.1)**

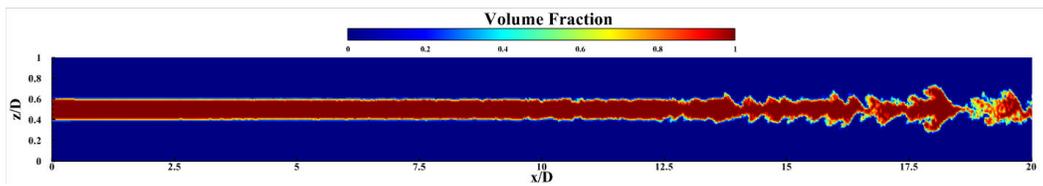

**(d) Distribution of volume fraction (ρ2/ρ1=1/10, H/D=0.2)**

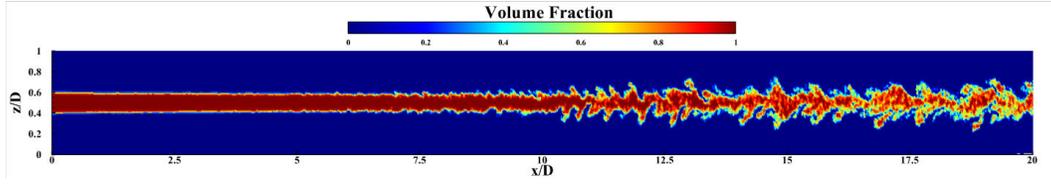

**(e) Distribution of volume fraction (ρ2/ρ1=1/100, H/D=0.2)**

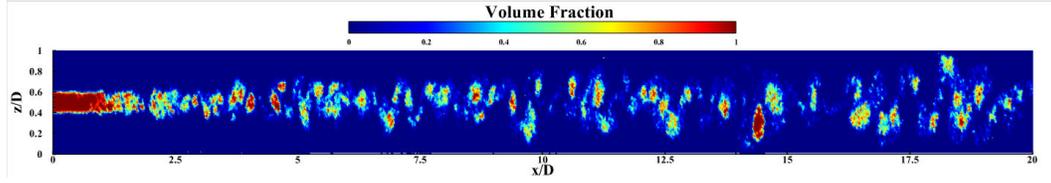

**(f) Distribution of volume fraction (ρ2/ρ1=1/1000, H/D=0.2)**

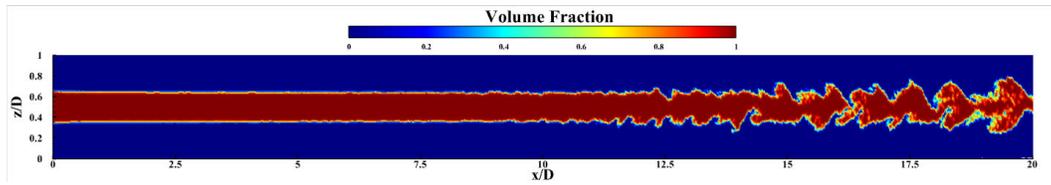

**(g) Distribution of volume fraction (ρ2/ρ1=1/10, H/D=0.3)**

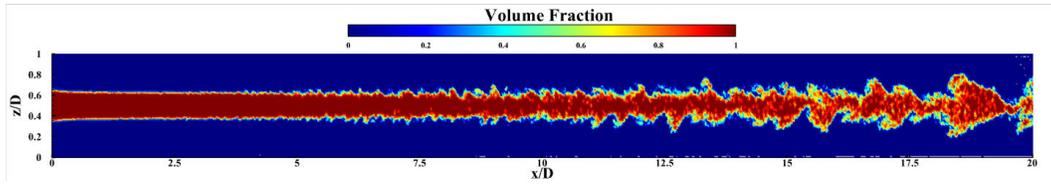

**(h) Distribution of volume fraction (ρ2/ρ1=1/100, H/D=0.3)**

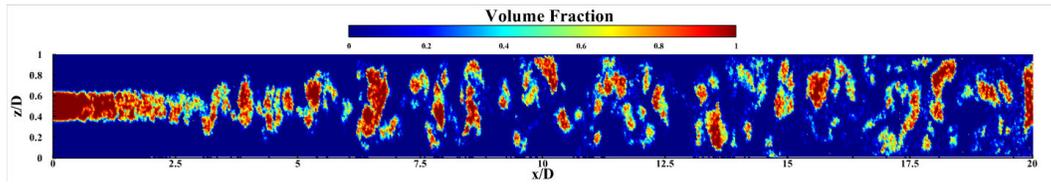

**(i) Distribution of volume fraction (ρ2/ρ1=1/1000, H/D=0.3)**

**Fig. 25. Numerical results of turbulent multiphase channel flow obtained by the present algorithm.**

The contours of the volume fraction with different density ratios are given in Fig. 25. It is shown that the two-phase interaction tends to be more intense as the density ratio increases, especially a significant distortion of the interface between the two phases can be observed in the case of a density ratio at 1000. This phenomenon can be explained by the Bernoulli effect, the pressure of the lighter phase reduces due to the greater velocity compared to the heavier phase, leading to interfacial deformation, which is amplified by large density ratios. Rational turbulent evolution can be observed along the channel flow, which demonstrates the capability of the proposed algorithm to recover the turbulent behaviors in multiphase channel flow.

## 5.2 Horizontal slug flow in the channel

The slug flow plays a significant role in several circumstances including the oil mixed transportation and the cooling processes for the reactor core, whose formation is often accompanied by violent pressure fluctuations and significant interface deformation, which induces potential hazards to the structure and poses great challenge to numerical simulations. To evaluate the robustness of the proposed method under the violent flow condition, the numerical test of a horizontal slug flow in the channel is carried out in this section. The experiment has been conducted by Höhne and Mehlhoop [4], which gives reliable reference to validate the proposed method.

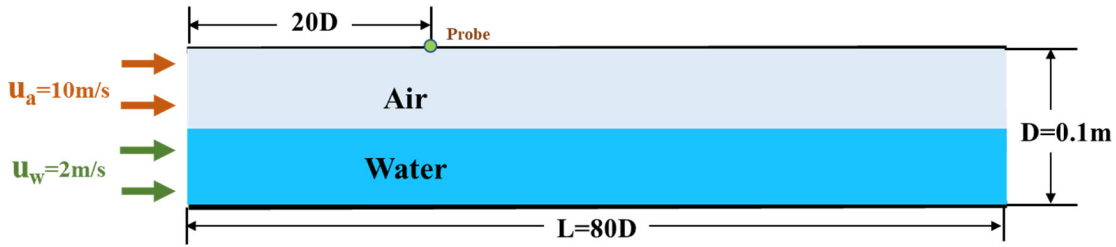

**Fig. 26. Schematic illustration of the horizontal slug flow in the channel.**

The detailed arrangement of the experiment is described in [4] and the schematic diagram of which is presented in Fig. 26. The diameter of the channel is D = 0.1m and the length is L=80D. The air and water enter the channel with constant velocity and a blade with 0.5m length is placed at the inlet to separate the two phases. The density of the air is $\rho_a$ = 1.204 kg/m$^3$ with a dynamic viscosity $\mu_a$ = 1.837×10$^{-5}$ Pa•s, and the physical properties of water are $\rho_w$ = 998.2 kg/m$^3$ and $\mu_w$ =

8.891×10$^{-4}$ Pa•s. The gravity effect is considered in this case with $g$ = 9.81m/s$^2$. Additionally, a pressure probe is set at the top of the channel to record the frequency of slug formation.

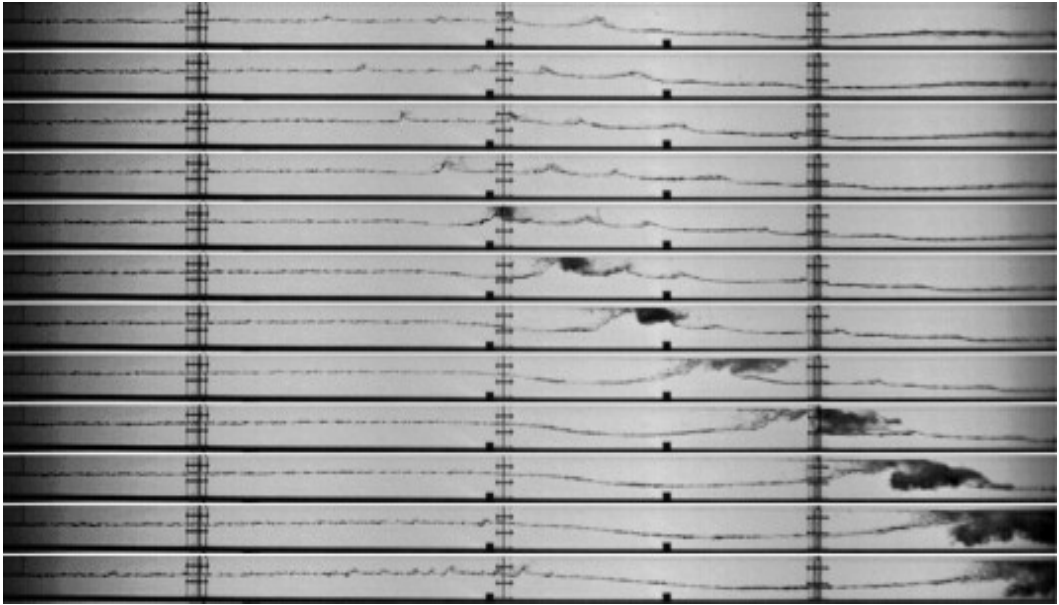

**Fig. 27. Experimental snapshots[4] of the flow patterns of the horizontal slug flow in the channel (0–3.2 m after the inlet)**

Evolutions of the flow pattern obtained by the present method are presented in Fig. 28 and compared with experimental results (Fig. 27) given in Ref.[4]. As it can be observed, the small-scale waves are generated by the two-phase interaction, and then the accumulation of which induces a bigger wave, which covers the downstream waves and consequently clogs the cross-section. Therefore, the overall evolution of the slug has been reproduced by the proposed method in comparison with the experiments, which gives initial validation of the accuracy of the proposed method.

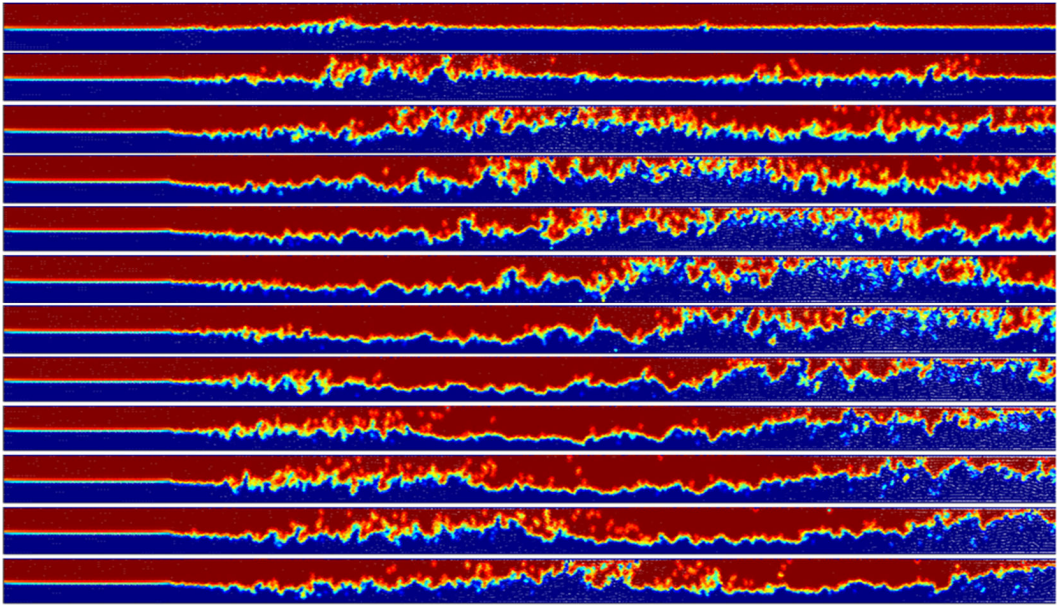

**Fig. 28. Numerical results of the flow patterns of the horizontal slug flow in the channel (0–3.2 m after the inlet)**

The evolution of pressure at the probe of the present SPH result is shown in Fig. 29, and three particle resolutions dx=D/10, dx=D/20, and dx=D/40 are adopted. It can be observed that the pressure amplitude shows a periodic rise and fall, corroborating that slug are generated at a certain frequency. To determine the characteristic frequency, the power spectral density of pressure is obtained through the fast Fourier transform (FFT), whose results are plotted in Fig. 30 with the labeling of characteristic slug frequency. With the increasing particle resolution, the characteristic slug frequency converges to the experimental value, which demonstrates the capability of the proposed method to reproduce the characteristic behavior of complex multi-phase channel flow.

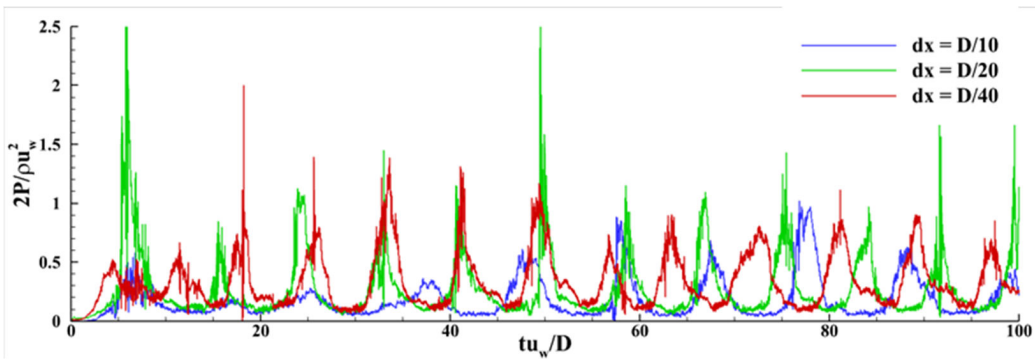

**Fig. 29. The evolution of pressure at the probe with different spatial resolutions of the horizontal slug flow in the channel.**

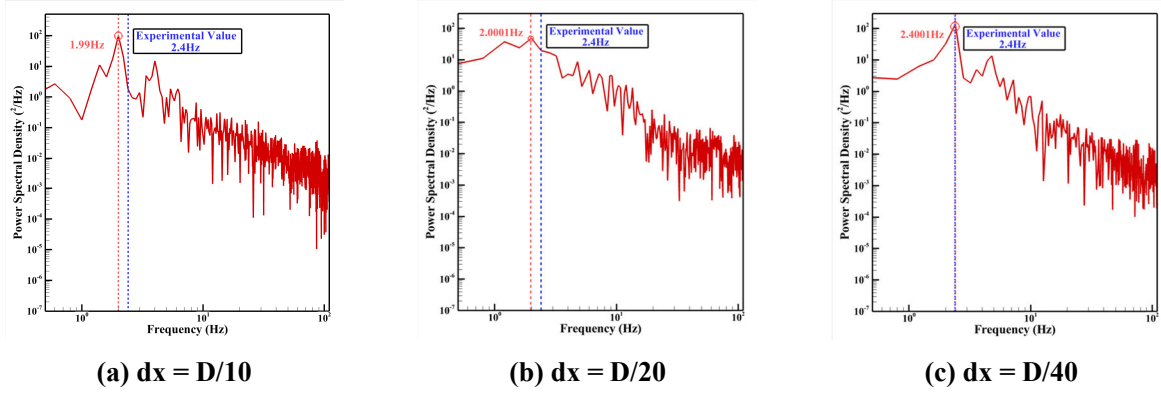

(a) dx = D/10　　　　　(b) dx = D/20　　　　　(c) dx = D/40

**Fig. 30. Power spectral density of pressure at the probe of the horizontal slug flow in the channel.**

## 6 Conclusions

In this study, a general SPH algorithm is developed for the effective simulation of complex multiphase flows with open boundaries. The behavior of particles is resolved by the WCSPH method with a general formulation for single-phase and multi-phase systems, and the SPH-based turbulent model is employed for handling complex flow scenarios in a more versatile manner.

To address numerical instabilities in extreme conditions, the open boundary condition is fine-tuned by two novel strategies. Firstly, a relaxation strategy is employed to achieve a stable inflow condition, where a density relaxation treatment is proposed to suppress the pressure oscillation of emitter particles with adaptability in single-phase and multi-phase scenarios, and the velocity relaxation is utilized for improved numerical stability in the buffer region. Secondly, a modified particle shifting technique (PST) is implemented to address the unphysical concentrations of particles, whose amplitude of correction in the interfacial region is adjusted by the color function. To guarantee a rational shifting intensity at the outlet boundary, an outflow limiter is implemented to locally adjust the particle shifting vector, which stabilizes the velocity evolution in the outflow region. Meanwhile, no additional extrapolated interpolation is required, which significantly saves the computational cost.

The accuracy and stability of the proposed algorithms are validated by a series of representative numerical examples with increasing complexity, including the single-phase Poiseuille flow, two-fluid Poiseuille flow, two-phase co-current flow, and the turbulent flow. Good agreements with the analytical solution or experimental data have been achieved in all examples, and the effectiveness of the novel strategies is confirmed through the evaluation. Further test of the algorithm's robustness is carried out through the multiphase channel flow in the turbulent state with variable density ratios, and rational evolutions of the turbulent behavior can be observed. Additionally, the horizontal slug channel flow case is conducted and performs a good reproduction of the experimental result, which confirms the capability of the present algorithms in recovering the characteristic behavior of complex channel flow. It is thus demonstrated that the proposed algorithm can give accurate, stable and robust solutions in complex multiphase flows with open boundary conditions.

## Appendix

Considering the time filtering treatment in Reynolds-Averaged Navier-Stokes Equations (RANS), the momentum equation can be rewritten as:

$$\frac{d\bar{\boldsymbol{u}}}{dt} = -\frac{\nabla p}{\rho} + \nu \nabla^2 \bar{\boldsymbol{u}} + \frac{\nabla \cdot \boldsymbol{\tau}}{\rho} \tag{41}$$

where $\boldsymbol{\tau}$ is the Reynolds stress tensor. According to the Boussinesq hypothesis, it can be expressed as:

$$\boldsymbol{\tau} = -\frac{2}{3}\rho k \boldsymbol{I} + 2\mu_t \boldsymbol{S} \tag{42}$$

Where $k$ is the turbulent kinetic energy, $\mu_t$ is the turbulent viscosity, and $\boldsymbol{S} = \frac{1}{2}[\nabla\overline{\boldsymbol{u}} + (\nabla\overline{\boldsymbol{u}})^T]$ is the time-averaged strain rate. To determine the above unknowns for the closure of governing equations, the two-equation k-ε model is used, thus, the turbulent viscosity can be calculated by[58]:

$$\mu_t = c_\mu \rho k / \varepsilon \tag{43}$$

Where $c_\mu$ is a constant with the value of 0.09, and $\varepsilon$ is the turbulent dissipation rate. Then, to solve the remaining unknowns $k$ and $\varepsilon$, the two specific equations are introduced[59]:

$$\rho \frac{dk}{dt} = \nabla \cdot \left[ \left( \mu + \frac{\mu_t}{\sigma_k} \right) \nabla k \right] + 2\mu_t \boldsymbol{S} : \boldsymbol{S} - \rho\varepsilon \tag{44}$$

$$\rho \frac{d\varepsilon}{dt} = \nabla \cdot \left[ \left( \mu + \frac{\mu_t}{\sigma_\varepsilon} \right) \nabla\varepsilon \right] + c_1 \frac{\varepsilon}{k} \boldsymbol{S} : \boldsymbol{S} - c_2 \rho \frac{\varepsilon^2}{k} \tag{45}$$

The constants in the above two equations are: $\sigma_k$=1.0, $\sigma_\varepsilon$=1.3, $c_1$=1.44, $c_2$=1.92.

## Acknowledgement